\documentstyle[12pt]{article}

\begin{document}

\author{Sandro S. Costa\\
Instituto de F\'\i sica Te\'orica -- IFT\\
Universidade Estadual Paulista -- UNESP\\
Rua Pamplona, 145 -- Bela Vista\\
01405-900 -- S\~ao Paulo -- SP\\
Brazil}
\title{Schwarzschild solution and Mach's principle}
\date{\today}
\maketitle

\begin{abstract}
The purpose of this work is to show a possible reconciliation between Mach's
principle and the Schwarzschild solution, using a description of this
solution in terms of a new radial coordinate related to the
behaviour of Unruh detectors.
\end{abstract}

\section{Introduction}

When beginning the study of General Relativity (see, for example, \cite
{Dinverno}) a student usually finds in his way two basic concepts:

\begin{itemize}
\item  the Mach's principle, that can be stated as \cite{Dinverno} {\bf %
``the matter distribution determines the geometry; if there is no matter
there is no geometry -- a body in an otherwise empty universe should possess
no inertial properties'';}

\item  the standard form of the Schwarzschild solution, 
\begin{equation}
\label{Traditionals}ds^2=\left( 1-\frac{2m}r\right) dt^2-\frac 1{\left( 1-%
\frac{2m}r\right) }dr^2-r^2d\Omega ^2
\end{equation}
where $d\Omega ^2\equiv \sin ^2\theta d\varphi ^2+d\theta ^2$.
\end{itemize}

However, these concepts seem to be disconnected. It is not clear if the
Mach's principle is obeyed or not in the Schwarzschild solution; even the
Minkowski solution seems to ignore the idea of `no matter, no geometry',
since its expression for the interval contains a spatial part completely
independent of matter.

The purpose of this work is, then, to obtain some kind of reconciliation
between Mach's principle and the classical solutions of General Relativity
(Kerr, Schwarzschild and Minkowski), using a description of this solutions
in terms of a new radial coordinate $\omega $, which, in the case of the
Kerr metric, is defined in the simple relation 
\begin{equation}
\label{radius0}r=2\cosh \frac \omega 2(m\cosh \frac \omega 2+a\sinh \frac
\omega 2)\;\; .
\end{equation}
For the special case of Schwarzschild solution this relation is simplified, 
\begin{equation}
\label{radius}r=2m\cosh ^2\frac \omega 2\;\; ,
\end{equation}
giving an interval with the spatial part of the metric directly related to
the existence of the mass: 
\begin{equation}
\label{Schwarzschild1}ds^2=\tanh ^2\frac \omega 2dt^2-4m^2\cosh ^4\frac
\omega 2\left[ d\omega ^2+d\Omega ^2\right] .
\end{equation}
In this case one possible physical interpretation of this coordinate, as
showed here, suggests that Unruh detectors -- theoretical devices used in
the context of Quantum Field Theory -- can play a crucial role in the
comprehension of the geometry of spacetime.

The work is organized as follows: the next section, {\bf `Mass and space'},
shows how to obtain the idea of `a space generated by mass' from an
expression of the Kerr interval and its correspondences with Minkowski and
Schwarzschild metrics. The section {\bf `Unruh detectors' }simply presents
the rate of excitation of an Unruh detector placed in the Schwarzschild
space and converts it to the new set of coordinates; it is in this
conversion that one can suggest an interpretation for the coordinates linked
with the behaviour of the detector. The last section, {\bf `Conclusions'},
presents some comments and ideas, discussing the connection between this
`new' representation of the Schwarzschild metric and the Mach's principle.

Through all this work natural units -- $\hbar =c=k=G=1$ -- are used; the
metric signature is $\left( +---\right) $.

\section{Mass and space}

\subsection{Horizon functions and Kerr metric}

One of the most generic solutions of General Relativity is the Kerr
solution, which has two parameters, $a$ and $m$, related respectively with
the rotation and mass of a black hole; this solution can be written as \cite
{Dinverno} 
$$
ds^2=\frac \Delta {\rho ^2}\left[ dt-a\sin ^2\theta d\varphi \right] ^2-%
\frac{\sin ^2\theta }{\rho ^2}\left[ \left( r^2+a^2\right) d\varphi
-adt\right] ^2-\rho ^2\left[ \frac{dr^2}\Delta +d\theta ^2\right]  
$$
where 
$$
\rho ^2=r^2+a^2\cos ^2\theta  
$$
and the `horizon function' \cite{Chandrasekhar}, $\Delta $, is 
\begin{equation}
\label{Delta3}\Delta =r^2-2mr+a^2\;\; .
\end{equation}
Using the coordinate transformation given by equation (\ref{radius0}) one has%
$$
\Delta =\left( m\sinh \omega +a\cosh \omega \right) ^2 
$$
and thus the expression for the interval becomes 
\begin{equation}
\label{Kerrinterval}ds^2=dt^2-\left[ a^2ds_a^2+m^2ds_m^2+ds_{am}^2\right] 
\end{equation}
where

\begin{equation}
ds_a^2=\left( \cosh ^2\omega -\sin ^2\theta \right) \left( d\omega
^2+d\theta ^2\right) +\cosh ^2\omega \sin ^2\theta d\varphi ^2 
\end{equation}

\begin{equation}
ds_m^2=4\cosh ^4\frac \omega 2\left[ d\omega ^2+d\Omega ^2\right] 
\end{equation}

\begin{eqnarray}
ds_{am}^2& = & 4m\cosh ^2\frac \omega 2\left[ a\sinh \omega \left( d\omega
^2+d\Omega ^2\right)\right.\nonumber \\ 
&& \left. +\frac{\left( dt-a\sin ^2\theta d\varphi \right)
^2\left( a\tanh \frac \omega 2+m\right) }{4\cosh ^2\frac \omega 2(m\cosh
\frac \omega 2+a\sinh \frac \omega 2)^2+a^2\cos ^2\theta }\right]  
\end{eqnarray}

and

$$
d\Omega ^2=d\theta ^2+\sin ^2\theta d\varphi ^2\;\; . 
$$
The connection between this solution and Minkowski and Schwarzschild metrics
is simple: when only $a=0$, one shall obtain Schwarzschild solution; when
both $a=0$ and $m=0$ one shall obtain Minkowski solution. These
correspondences will be seen next.

\subsection{Minkowski metric}

When $m=0$ expression (\ref{Kerrinterval}) becomes
\begin{equation}
\label{Kerr1}ds^2=dt^2-a^2\left[ \left( \sinh ^2\omega +\cos ^2\theta
\right) \left( d\omega ^2+d\theta ^2\right) -\cosh ^2\omega \sin ^2\theta
d\varphi ^2\right] . 
\end{equation}
This expression is easily obtained from the standard Minkowski metric: the 
Minkowski metric is, in spherical coordinates, 
\begin{equation}
\label{spherical}ds^2=dt^2-\left[ dr^2+r^2\left( \sin ^2\theta d\varphi
^2+d\theta ^2\right) \right] . 
\end{equation}

This expression for the interval of Minkowski space can be seen as a
particular case of an oblate ellipsoidal metric \cite{Landau}, which
contains an asymmetry in the plane $xy$: 
\begin{equation}
\label{co}
\begin{array}{c}
x=\left( r^2+a^2\right) ^{\frac 12}\sin \theta \cos \varphi =\Delta ^{\frac
12}\sin \theta \cos \varphi  \\ 
y=\left( r^2+a^2\right) ^{\frac 12}\sin \theta \sin \varphi =\Delta ^{\frac
12}\sin \theta \sin \varphi  \\ 
z=r\cos \theta 
\end{array}
\end{equation}
{\it i.e.}, 
\begin{equation}
\label{Kerr}ds^2=\left[ dt-\Delta ^{\frac 12}\sin \theta d\varphi \right]
\left[ dt-\Delta ^{\frac 12}\sin \theta d\varphi \right] -\left( r^2+a^2\cos
^2\theta \right) \left[ \frac{dr^2}\Delta +d\theta ^2\right] 
\end{equation}
where 
\begin{equation}
\label{oblate}\Delta \equiv r^2+a^2\;\; .
\end{equation}
In this case the surfaces with constant $r$ are oblate ellipsoids of
rotation, obeying the equation 
\begin{equation}
\label{elipsoidal}\frac{x^2+y^2}\Delta +\frac{z^2}{r^2}=1
\end{equation}
It is easy to see that if one chooses a change of coordinates such that  
\begin{equation}
\label{cosh}\Delta =a^2\cosh ^2\omega 
\end{equation}
one can rewrite the metric as in expression (\ref{Kerr1}); in this
expression it is clear that when the parameter $a$, responsible for the
perturbation in the plane $xy$, vanishes the metric becomes dependent only
on the time coordinate. This is a result which follows the idea that the
space is generated by perturbations in the vacuum; almost the same procedure
can be used to obtain the solution for a perturbation of the vacuum by a
spherical distribution of mass.

\subsection{Schwarzschild metric: external solution}

When $a=0$ expression (\ref{Kerrinterval}) becomes 
$$
ds^2=\tanh ^2\frac \omega 2dt^2-4m^2\cosh ^4\frac \omega 2\left[ d\omega
^2+d\Omega ^2\right] . 
$$
The connection between this one and the standard form of Schwarzschild
solution is also easy: the standard Schwarzschild metric can be written as 
\begin{equation}
\label{Schwarzschild0}ds^2=\frac \Delta {r^2}\left[ dt-\frac{r^2\sin \theta
d\varphi }{\Delta ^{\frac 12}}\right] \left[ dt+\frac{r^2\sin \theta
d\varphi }{\Delta ^{\frac 12}}\right] -r^2\left[ d\theta ^2+\frac{dr^2}%
\Delta \right] 
\end{equation}
where we use a new definition for the function $\Delta $, containing only a
term linearly proportional to the radius, 
\begin{equation}
\label{Delta2}\Delta \equiv r^2-2mr\;\; .
\end{equation}
One can choose now the coordinate transformation 
\begin{equation}
\label{sinh}\Delta =m^2\sinh ^2\omega 
\end{equation}
-- which, in turn, implies in the expression (\ref{radius}) 
$$
r=2m\cosh ^2\frac \omega 2 
$$
-- obtaining 
$$
ds^2=\tanh ^2\frac \omega 2dt^2-4m^2\cosh ^4\frac \omega 2\left[ d\omega
^2+d\Omega ^2\right] 
$$
where it was used that $\sinh ^2\omega =4\cosh ^2\frac \omega 2\sinh ^2\frac
\omega 2$, and $d\Omega ^2\equiv \sin ^2\theta d\varphi ^2+d\theta ^2$.

This expression is the solution presented in the introduction as equation (%
\ref{Schwarzschild1}); it shall be noticed that this solution is valid only
for $r\geq 2m$, and it presents only one apparent singularity: when $\omega
\rightarrow \infty $, the spatial part of the metric `explodes'; but this is
a problem of the coordinates, since the curvature scalar is everywhere null,
including the points where $r=2m$.

\subsection{Schwarzschild metric: internal solution}

In order to extend the range of the solution given in expression (\ref
{Schwarzschild1}) one can extend the domain of the angular coordinate $%
\omega $ into the set of complex numbers: $\omega \rightarrow \omega
+i\Sigma $. Then, the adequate coordinate transformation turns out to be 
\begin{equation}
\label{complex}r=2m\left| \cosh \left( \frac{\omega +i\Sigma }2\right)
\right| ^2=2m\left[ \left( \cosh \frac \omega 2\cos \frac \Sigma 2\right)
^2+\left( \sinh \frac \omega 2\sin \frac \Sigma 2\right) ^2\right] .
\end{equation}
In this case the range of the coordinates is such that outside the surface $%
r=2m$, {\it i.e.}, for $r\geq 2m$, $\Sigma =0$ and $0\leq \omega <\infty $;
and for $r\leq 2m$, $\omega =0$ and $0\leq \Sigma \leq 2\pi $.

Though this use of a system with complex coordinates can be seen as an
artificial procedure, it has a simple 3-dimensional analogue: a system
constituted by a 3-dimensional hemisphere `glued' by its circular border in a
plane that has a circle removed; 2-dimensional beings walking in this
surface can use a system of coordinates that describes their distances from
the center of the hemisphere -- that always changes for paths {\it in the
plane} with a radial component -- and another system of coordinates to
describe the movements inside the hemisphere, where the distance from the
center of the hemisphere is always constant. Therefore, assuming equation (%
\ref{complex}) one can write in one expression the solution for the two
regions of the space,%
\begin{eqnarray}
  ds^2&=&\frac{\sinh ^2\frac \omega 2-\sin ^2\frac \Sigma 2}{\cosh ^2%
  \frac \omega 2-\sin ^2\frac \Sigma 2}dt^2-4m^2\left| \cosh%
  \left( \frac{\omega +i\Sigma }2\right) \right| ^4\nonumber \\
   && \times\left\{ \frac{\left( \sinh \frac \omega 2\cosh%
   \frac \omega 2d\omega -\sin \frac \Sigma 2\cos%
   \frac \Sigma 2d\Sigma \right) ^2}%
   {\left( \cosh ^2\frac \omega 2-\sin ^2\frac \Sigma 2\right)%
   \left( \sinh ^2\frac \omega 2-\sin ^2\frac \Sigma 2\right) }+%
   d\Omega ^2\right\} 
\end{eqnarray}
always observing the right domain of the coordinates: the space, though
embedded in a 5-dimensional frame, is {\it everywhere} 4-dimensional.

\section{Unruh detectors}

An Unruh detector is a two-level device that can be characterized by its
energy gap $E$; this detector, when immerse in a bath of particles, acts as
a filter, counting only the particles that possess the exact energy to fit in
its gap. Such detector, when in rest in a bath of scalar particles of
temperature $T$, gives a response proportional to the Planckian
distribution, 
\begin{equation}
\label{Planckian}n=\frac 1{e^{E/T}-1}\;\; , 
\end{equation}
as one should expect. However, in Minkowski vacuum, an Unruh detector with
constant acceleration $a$ also sees a Planckian distribution of particles,
corresponding to a thermal bath with temperature $\frac a{2\pi }$; this
strange -- and quite unexpected -- behaviour of the detector can be
explained by the idea that the particles seen by the detector are not
`real': they are particles created by the energy supplied to accelerate the
detector\footnote{%
For a full explanation about Unruh detectors see, for example, \cite{BD}.}.

In the Schwarzschild vacuum it can be shown -- see \cite{BD} or \cite{Hinton}
-- that an Unruh detector, fixed in a point around the central massive body,
also gives a response proportional to a Planckian distribution of particles
of temperature 
\begin{equation}
\label{temperature}T=\left[ 64\pi ^2M^2\left( 1-\frac{2M}r\right) \right]
^{-\frac 12} .
\end{equation}
Notice that this temperature never goes to zero, even when $r\rightarrow
\infty $. In terms of the angular coordinate $\omega $ this can be written
as 
\begin{equation}
\label{temperature1}T=\frac \kappa {2\pi }\coth \frac \omega 2
\end{equation}
where 
\begin{equation}
\label{supf}\kappa =\frac 1{4M}
\end{equation}
is the superficial gravity associated with a black hole.

From this last result, equation (\ref{temperature1}), one can see an easy
interpretation for the radial coordinate $\omega $: it is a coordinate
analogous to the angle $\theta $, defined as {\it rapidity} in the context
of Special Relativity, since\footnote{Here the constants $c$ and $k$ appear 
just to remind the usual definitions of $\beta $, 
both in Special Relativity and Thermodynamics.}%
$$
\beta _{\left( relativity\right) }\equiv \frac vc\equiv \tanh \theta  
$$
and, in this case,%
$$
\beta _{\left( thermodynamics\right) }\equiv \frac 1{kT}\equiv \frac{2\pi }%
\kappa \tanh \frac \omega 2 \;\; . 
$$
Thus, the coordinate $\omega $ reescales the Schwarzschild metric in
function of the temperature of the black hole.

\section{Conclusions}

The proposal of this work was to show an `alternative' interpretation of the
Schwarzschild solution, beginning by a description of this solution in
terms of a set of coordinates that contains an `angular radial' coordinate $%
\omega $; the physical meaning of such coordinate can be seen in the study
of the behaviour of an Unruh detector placed in a fixed point around the
spherical mass distribution, associated with a damping of the energies of
the scalar particles `seen' by the detector (since $\left| E\tanh \frac
\omega 2\right| \leq E$).

Following this description one can suggest that mass generates the space,
in a sense that in vacuum there must be no other distinguishable direction
than the direction of time; to put it in another -- very heterodox -- way,
one can suggest that space is a consequence of the presence of a mass 
({\it i.e.}, is the vacuum a physical expression of the gravitational 
field or simply the scenery where gravity plays?).

Surely, this is not the only interpretation of the expression (\ref
{Schwarzschild1}) 
$$
ds^2=\tanh ^2\frac \omega 2dt^2-4m^2\cosh ^4\frac \omega 2\left[ d\omega
^2+d\Omega ^2\right] 
$$
for the interval of Schwarzschild space: one can say that all that was done
was to use mass as a simple geometric parameter, exactly as it is done in
the obtention of the isotropic form of the Schwarzschild solution\footnote{%
The relation between the common spherical and the isotropic coordinates is $%
r=\rho \left( 1+\frac m{2\rho }\right) ^2$.},%
$$
ds^2=\left[ \frac{1-\frac m{2\rho }}{1+\frac m{2\rho }}\right] ^2dt^2-\left(
1+\frac m{2\rho }\right) ^4\left[ d\rho ^2+\rho ^2\left( \sin ^2\theta
d\varphi ^2+d\theta ^2\right) \right] 
$$
in order to obtain a solution that reminds the Euclidean 3-space for
hipersurfaces of constant time.

The approach here presented, however, even if results in a expression for
the interval similar to the isotropic form, tries prioritary to emphasize
the importance of the time coordinate in {\it any} space-time that is empty,
what recalls the Mach's principle: {\it there is no meaning in the idea of
absolute motion}; the expression (\ref{Schwarzschild1}) just states the
importance of the proper time -- and {\it it is the proper time that is
essential in the understanding of the behaviour of an Unruh detector}.

Another observation to be done is that the apparent explosion of the spatial
part of the metric present in this expression when $\omega \rightarrow
\infty $ can be avoided only if one makes $m\rightarrow 0$; this can be seen
as expressing the idea that the influence of mass is infinite in distance,
and thus no one can go to an infinite `angular distance' from it. Again this
is a restatement of the Mach principle.

To finish, it is worth to note that the expression (\ref{Kerr1}) for the
vacuum interval with anisotropy%
$$
ds^2=dt^2-a^2\left[ \left( \sinh ^2\omega +\cos ^2\theta \right) \left(
d\omega ^2+d\theta ^2\right) -\cosh ^2\omega \sin ^2\theta d\varphi
^2\right]  
$$
can also be seen as a representation of the Mach's principle, if one
associates a rotation with the parameter $a$; this rotation -- a rotation of
a particle without mass -- is a kind of absolute movement that can be
detected, since it generates a centrifugal acceleration\footnote{%
In \cite{Harrison} appears a very clear and simple discussion of the Mach's
principle that, as a matter of fact, served as a `seed' for this work.}.

\section*{Acknowledgements}
This work was completely supported by ``Funda\c c\~ao de Amparo \`a Pesquisa 
do Estado de S\~ao Paulo'' -- FAPESP. The author also must thanks to Prof. 
Roberto A. Kraenkel and to Armando Bernui, who, among other people in 
IFT, CBPF, UFRN and UFPB, made possible the participation of the author 
in a workshop -- ``A Estrutura Topol\'ogica do Universo'' (The Topological 
Structure of the Universe) -- in Natal, Brazil, where this work finally 
became mature.

\end{document}